\documentclass[aps,twocolumn,showpacs]{revtex4}
\usepackage{mathrsfs}
\begin{document}
\title{Spin and orbital angular momentum in gauge theories (I):
QED and determination of the angular momentum density}
\author{Xiang-Song Chen\footnote{Email: cxs@scu.edu.cn}, Xiao-Fu L\"{u}}
\affiliation{Department of Physics, Sichuan University, Chengdu
610064, China}
\author{Wei-Min Sun, Fan Wang}
\affiliation{Department of Physics, Nanjing University, Nanjing
210093, China}
\author{T. Goldman}
\affiliation{Theoretical Division, Los Alamos National Laboratory,
Los Alamos, NM 87545, USA}
\date{\today}

\begin{abstract}
This two-paper series addresses and fixes the long-standing gauge
invariance problem of angular momentum in gauge theories. This QED
part reveals: 1) The spin and orbital angular momenta of electrons
and photons can all be consistently defined gauge invariantly. 2)
These gauge-invariant quantities can be conveniently computed via
the canonical, gauge-dependent operators (e.g, $\psi ^\dagger \vec x
\times\frac 1i \vec \nabla \psi$) in the Coulomb gauge, which is in
fact what people (unconsciously) do in atomic physics. 3) The
renowned formula $\vec x\times\left(\vec E\times\vec B\right)$ is a
wrong density for the electromagnetic angular momentum. The angular
distribution of angular-momentum flow in polarized atomic radiation
is properly described not by this formula, but by the gauge
invariant quantities defined here. The QCD paper \cite{Chen07} will
give a non-trivial generalization to non-Abelian gauge theories, and
discuss the connection to nucleon spin structure.

\end{abstract}
\maketitle

Energy, momentum, and angular momentum are among the most important
properties of a physical field, in both classical and quantum
theories. It should be emphasized that these quantities are not
uniquely determined by the field Lagrangian. Their definitions allow
for certain arbitrariness, which must be fixed by other physical
requirements or by experiments. Gauge invariance is one such
requirement. It helps to fix the form of the energy-momentum tensor,
which is to be coupled to the gravitational field, therefore must
have a gauge invariant {\em density}. Certainly the gauge invariance
criteria also applies to the angular momentum. But up to now it
seemed to bring more uneasy feelings rather than assistance. For one
example, the labeling of atomic states employs the electron orbital
angular momentum operator $\vec L_e=\int d^3x \psi ^\dagger \vec x
\times\frac 1i \vec \nabla \psi$, but this operator is gauge
dependent! For another example, it was taught in common textbooks
that gauge invariance prohibits the separation of photon angular
momentum into spin and orbital contributions. \cite{Jauc55,Bere82}

The goal of this paper is to turn the situation around. Our strategy
is to investigate the spin and orbital angular momenta of electrons
and photons in a whole QED system, and examine the angular momentum
at the same footing as the energy-momentum, namely, by seeking an
appropriate density expression. We will see that examination of the
standard angular distribution in polarized atomic radiation can
actually do the job.

We start with the ``canonical'' angular momentum operators in QED:
\begin{eqnarray}
\vec J_{QED} &=& \int d^3 x \psi ^\dagger \frac 12 \vec \Sigma \psi
+ \int d^3x \psi ^\dagger \vec x \times\frac 1i \vec \nabla \psi
\nonumber \\
 &+&\int d^3x \vec E\times \vec A+
\int d^3x E^i\vec x\times \vec \nabla A^i \nonumber \\
&\equiv& \vec S_e +\vec L_e +\vec S_\gamma +\vec L_\gamma.
\label{J1}
\end{eqnarray}
These operators are termed ``canonical'' ones because their
individual roles in generating spatial rotations can be directly
recognized. These expressions are derived straightforwardly by
applying N\"other's theorem to the QED Lagrangian
\begin{equation}
\mathscr{L}=-\frac 14 F_{\mu\nu}F^{\mu\nu}+ \bar\psi\left(
i\gamma^\mu D_\mu -m \right)\psi. \label{L}
\end{equation}

The canonical angular momentum operators in Eq. (\ref{J1}) are what
people use familiarly in discussing polarized atomic states and
radiations. However, except for the electron spin, all other three
terms are gauge dependent. This brings a very uneasy concern about
the physical meanings of these quantities, and even the whole
labeling of atomic states.

Before the above subtlety was clearly up, its counterpart in QCD was
sharply faced when people tried to understand the nucleon spin in
terms of the spin and orbital contributions of quarks and gluons.
\cite{EMC} To satisfy the gauge invariance requirement, people
considered an alternative, explicitly gauge invariant decomposition
of the QCD angular momentum. \cite{Ji97,Chen97} Its QED version is
\begin{eqnarray}
\vec J_{QED} &=& \int d^3 x \psi ^\dagger \frac 12 \vec \Sigma \psi
+ \int d^3x \psi ^\dagger \vec x \times\frac 1i \vec D
\psi\nonumber\\
&+&\int d^3x \vec x\times \left(\vec E\times\vec B\right) \nonumber \\
&\equiv& \vec S_e +\vec L'_e +\vec J'_\gamma.  \label{J2}
\end{eqnarray}

This is obtained from Eq. (\ref{J1}) by adding a surface term which
vanishes after integration. At first sight, this decomposition might
appear satisfactory, because $\vec x\times \left(\vec E\times\vec
B\right)$ is the familiar electromagnetic angular momentum
constructed with the renowned Poynting vector. However, the gauge
invariance of Eq. (\ref{J2}) does not help to justify the labeling
of atomic states, which uses eigenvalues of the canonical $\vec
L_e$, not the gauge invariant $\vec L'_e$. It does not answer either
the physical meanings of the photon spin and orbital angular
momentum, especially the measurements for them: The photon spin has
been measured directly by Beth over 70 years ago. \cite{Beth36}
Recently, detection and manipulation of the photon orbital angular
momentum have also been carried out, and became a hot topic due to
its potential application in quantum information processing.
\cite{Enk07,Ande06,Alex06,Marr06,Leac02,Alle92}

As we advocated at the beginning, we seek to remove the
arbitrariness in defining angular momentum by examining the density.
When brought to this trial, Eqs. (\ref{J1}) and (\ref{J2}) cannot
possibly both survive, because they give different angular momentum
densities, although the integrated $\vec J_{QED}$ is the same.
Unlike the energy-momentum tensor, the rank-3 angular momentum
tensor does not couple to any physical field, so no theoretical clue
can be found in this regard. We have to ask what kind of experiments
can measure the angular momentum density. The answer we provide here
is the standard angular distribution in polarized atomic radiations.

Let us consider a pure electric multipole radiation of order
$(l,m)$. (The discussion for magnetic-type radiation is exactly
similar.) The photon wave-function is given by
\begin{eqnarray}
\vec B_{lm}&=&a_{lm} j_l(kr) \vec L Y_{lm}\nonumber \\
\vec E_{lm}&=&i k \vec A_{lm} =\frac ik \vec \nabla \times \vec
B_{lm}. \label{wave}
\end{eqnarray}
Here $\vec A_{lm}$ is written in the Coulomb gauge. $\vec L\equiv
\vec x\times \frac 1i \vec \nabla$. The wave amplitude $a_{lm}$ is
related to the emission probability, which in turn is determined by
the transition matrix element of the electric $2^l$-pole moment. The
angular distribution of this radiation is
\begin{equation}
\frac{dP}{d\Omega}=\frac{1}{2k^2}\cdot\frac{1}{l(l+1)}
\left|a_{lm}\right|^2 \left|\vec L Y_{lm} \right|^2. \label{angu}
\end{equation}
The flow of angular momentum must also follow this distribution,
because an emitted photon takes off all the energy and angular
momentum lost by the polarized source atom. Now we are at the
position to check which definition of angular momentum density can
produce the above angular distribution.

Before promptly discarding the gauge-dependent density given in Eq.
(\ref{J1}), a moment's thought tells that it can actually agree
perfectly with Eq. (\ref{angu}), but only in the Coulomb gauge, in
which the photon wave-function $\vec A_{lm}$ in Eq. (\ref{wave}) is
the eigenstate of $(J_\gamma)_ z=(S_\gamma)_z+(L_\gamma)_z$.

Then, how about the gauge invariant $\vec x\times\left(\vec
E\times\vec B\right)$. Awkwardly, this renowned formula performs
badly. Take the simplest electric dipole radiation as an example.
Setting $(l,m)=(1,1)$ in Eq. (\ref{angu}) gives
\begin{equation}
\frac{dP}{d\Omega}=k\frac{dJ_z}{d\Omega}=\frac{1}{2k^2}
\cdot\frac{3}{8\pi}\left|a_{lm}\right|^2
\cdot \frac 12 \left(1+\cos ^2\theta\right). \label{11}
\end{equation}
However, $\vec x\times\left(\vec E\times\vec B\right)$ leads to a
flow of angular momentum projection $J_z$ according to
\begin{equation}
\frac{dJ_z}{d\Omega}=\frac{1}{2k^3}\cdot\frac{3}{8\pi}
\left|a_{lm}\right|^2\sin^2\theta. \label{11'}
\end{equation}
This strongly disagree with Eq. (\ref{11}), and is evidently wrong:
In electric $(1,1)$ radiation each emitted photon takes off $1\hbar$
of $J_z$, but Eq. (\ref{11'}) says that no $J_z$ flows along the $z$
axis, whereas along the $x$-$y$ plane one emitted photon should
carry $2\hbar$ of $J_z$.

In fact, there had been various clear hints that $\vec
x\times\left(\vec E\times\vec B\right)$ is not a correct description
of the electromagnetic angular momentum density. (The most
well-known example is probably the plane-wave paradox.) In
\cite{Chen97}, two of us (Chen and Wang) pointed out that when the
electromagnetic field interacts with the Dirac field,
$J'_\gamma=\int d^3 x \vec x\times\left(\vec E\times\vec B\right)$
is not its rotation generator. This can be easily seen from Eq.
(\ref{J2}): The total $\vec J_{QED}$ and the electron spin $\int d^3
x \psi ^\dagger \frac 12 \vec \Sigma \psi $ both satisfy the angular
momentum algebra $\vec J\times\vec J=i\vec J$, but since this
algebra is obviously violated by the operator $\vec L'_e= \int d^3x
\psi ^\dagger \vec x \times\frac 1i \vec D \psi$, it must also be
violated by the remaining term $J'_\gamma$, which therefore cannot
possibly be a rotation generator.

Now that we have ruled out the explicitly gauge-invariant Eq.
(\ref{J2}), while Eq. (\ref{J1}) can only do a good job in a
specific (the Coulomb) gauge, where is the satisfactory solution? To
a large extent, the solution had actually been found over ten year
ago. \cite{Enk94} (See also a recent discussion in \cite{Calv06}.)
When organized consistently, the gauge invariant, physically
reasonable expression for QED angular momentum reads
\begin{eqnarray}
\vec J_{QED} &=& \int d^3 x \psi ^\dagger \frac 12 \vec \Sigma \psi
+ \int d^3x \psi ^\dagger \vec x \times\frac 1i \vec D_{pure} \psi
\nonumber \\
 &+&\int d^3x \vec E \times \vec A_{phys}+ \int d^3x
E^i\vec
x\times \vec \nabla A_{phys}^i \nonumber \\
&\equiv& \vec S_e +\vec L''_e +\vec S''_\gamma +\vec L''_\gamma
\label{J3}.
\end{eqnarray}

Here, $\vec D_{pure} \equiv\vec \nabla -ie\vec A_{pure}$, $\vec
A_{pure}+\vec A_{phys}\equiv \vec A$ are defined through:
\begin{eqnarray}
\vec \nabla\cdot \vec A_{phys} =0,\label{A2}\\
\vec \nabla \times \vec A_{pure}=0. \label{A3}
\end{eqnarray}
They are nothing but the transverse and longitudinal components of
the vector potential $\vec A$. The suffixes we use here is to make
their physical contents clear, and to make preparation for
generalizations to QCD \cite{Chen07}. With the boundary condition
that $\vec A$, $\vec A_{pure}$, and $\vec A_{phys}$ all vanish at
spatial infinity, Eqs. (\ref{A2}) and (\ref{A3}) prescribe a unique
decomposition of $\vec A$ into $\vec A_{pure}$ and $\vec A_{phys}$,
and dictates their gauge transformation properties:
\begin{eqnarray}
\vec A_{pure}&\rightarrow & \vec A'_{pure}= \vec A_{pure} +\vec
\nabla \Lambda, \label{Apure}\\
\vec A_{phys}&\rightarrow &\vec A'_{phys}= \vec A_{phys}.
\label{Aphys}
\end{eqnarray}
Eqs. (\ref{A3}) and (\ref{Apure}) tell that in QED $\vec A_{pure}$
is a pure gauge field in all gauges, and it transforms in the same
manner as the full vector field $\vec A$ does:
\begin{equation}
\vec A\rightarrow \vec A'= \vec A +\vec \nabla \Lambda.
\end{equation}
On the other hand, the transverse field $\vec A_{phys}$ is
unaffected by gauge transformation, thus can be regarded as the
``physical'' part of $\vec A$.

Now we have all the need elements to explain that Eq. (\ref{J3})
gives the correct expressions of spin and orbital angular momenta of
electrons and photons, including their densities. First of all, the
total $\vec J_{QED}$ given by Eq. (\ref{J3}) equals that in Eqs.
(\ref{J1}) and (\ref{J2}). This can be proved by writing $\vec
B=\vec \nabla \times \vec A_{phys}$ in Eq. (\ref{J2}), performing an
integration by parts, and rearranging the results. Secondly, the
gauge transformation properties of $\vec A_{pure}$ and $\vec
A_{phys}$ tell that each {\it density} term in Eq. (\ref{J3}) is
separately gauge invariant (certainly so is the integrated
operator). Thirdly, like the canonical $\vec L_e$, the gauge
invariant $\vec L''_e$ satisfies the algebra $\vec J\times\vec J=i
\vec J$. This is due to the property of $\vec A_{pure}$ in Eq.
(\ref{A3}). And finally, we note that in the Coulomb gauge $\vec
\nabla \cdot \vec A=0$, the longitudinal (pure-gauge) field $\vec
A_{pure}$ vanishes, thus all quantities in Eq. (\ref{J3}) coincide
with their canonical counterparts in Eq. (\ref{J1}). This
observation is of vital importance. It reveals that the gauge
invariant quantities in Eq. (\ref{J3}) can all be conveniently
computed via the canonical operators in the Coulomb gauge. This is
actually what people (unconsciously) do in studying atomic and
electromagnetic angular momenta, including the recent measurements
of the photon orbital angular momentum
\cite{Enk07,Ande06,Alex06,Marr06,Leac02,Alle92}. It is thus
understandable why these studies always get reasonable results.

After confirming that Eq. (\ref{J3}) is indeed the correct and
satisfactory answer for angular momenta in QED, one may feel like to
talk some ``latter-wit'' about it: The form of Eq. (\ref{J3}) could
have been guessed out by reasonable physical considerations: The
photon angular momentum should contain only the ``physical'' part of
the gauge field, which nevertheless should not appear in the orbital
angular momentum of the electron. The latter should thus only
include the non-physical $\vec A_{pure}$ to cancel the also
non-physical phase dependence of the electron field, keeping the
whole $\vec L''_e$ gauge invariant. Honestly, these ``physical
considerations'' hardly helped us in writing down Eq. (\ref{J3}),
but such ``latter-wit'' for QED did serve as an important guidance
to a non-trivial solution for the angular momentum in QCD
\cite{Chen07}.

The authors acknowledge a helpful discussion with C.D. Roberts. This
research is supported in part by the China National Science
Foundation under grants 10475057 and 90503011, and in part by the US
Department of Energy under contract W-7405-ENG-36.

\end{document}